\documentclass[twocolumn,showpacs,preprintnumbers,superscriptaddress,amsmath,floatfix,amssymb,secnumarabic]{revtex4}

\usepackage[colorlinks=true]{hyperref}
\usepackage{graphicx}

\newcommand{\comment}[1]{}

\newcommand{\nm}{\, {\rm nm}}
\newcommand{\ev}{\, {\rm eV}}
\newcommand{\K}{\, {\rm K}}

\newcommand{\lr}[1]{ \left( #1 \right) }
\newcommand{\lrs}[1]{ \left[ #1 \right] }

\newcommand{\vev}[1]{ \langle \, #1 \, \rangle }

\newcommand{\tr}{ {\rm Tr} \, }

\newcommand{\ket}[1]{ \, | #1 \rangle }
\newcommand{\bra}[1]{ \langle #1 | \, }

\renewcommand{\det}[1]{ {\rm det} \left( #1 \right) }

\newcommand{\expa}[1]{ \exp{\left( #1 \right)} }

\newcommand{\logo}{\\ \vskip -18mm
\leftline{\includegraphics[scale=0.3,clip=false]{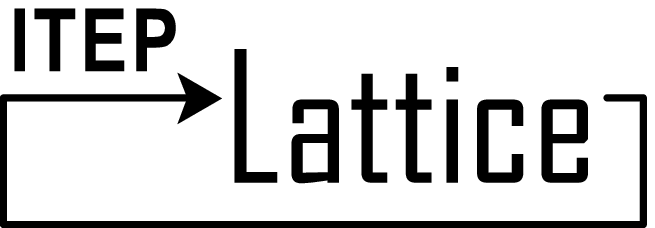}} \vskip 10mm}

\begin{document}
\sloppy

\title{Monte-Carlo study of the semimetal-insulator phase transition in monolayer graphene with realistic inter-electron interaction potential\logo}

\author{M.V. Ulybyshev}
\email{ulybyshev@goa.bog.msu.ru}
\affiliation{ITEP, B. Cheremushkinskaya str. 25, Moscow, 117218 Russia}
\affiliation{Institute for Theoretical Problems of Microphysics, Moscow State University, Moscow, 119899 Russia}

\author{P. V. Buividovich}
\email{Pavel.Buividovich@physik.uni-regensburg.de}
\affiliation{Institut f\"{u}r Theoretische Physik, Universit\"{a}t Regensburg, D-93040 Regensburg, Germany}

\author{M. I. Katsnelson}
\email{M.Katsnelson@science.ru.nl}
 \affiliation{Radboud University Nijmegen, Institute for Molecules and Materials, Heyndaalseweg 135, NL-6525AJ Nijmegen, The Netherlands}

\author{M. I. Polikarpov}
\email{polykarp@itep.ru}
 \affiliation{ITEP, B. Cheremushkinskaya str. 25, Moscow, 117218 Russia}
 \affiliation{Moscow Inst Phys \& Technol, Institutskii per. 9, Dolgoprudny, Moscow Region, 141700 Russia}

 \date{June 29, 2013}

\begin{abstract}
 We report on the results of the first-principle numerical study of spontaneous breaking of chiral (sublattice) symmetry in suspended monolayer graphene due to electrostatic interaction, which takes into account the screening of Coulomb potential by electrons on $\sigma$-orbitals. In contrast to the results of previous numerical simulations with unscreened potential, we find that suspended graphene is in the conducting phase with unbroken chiral symmetry. This finding is in agreement with recent experimental results by the Manchester group \cite{Elias:11:1}. Further, by artificially increasing the interaction strength we demonstrate that suspended graphene is quite close to the phase transition associated with spontaneous chiral symmetry breaking, which suggests that fluctuations of chirality and nonperturbative effects might still be quite important.
\end{abstract}
\maketitle

 In recent years significant effort has been invested into numerical studies of the electronic transport properties of ideal monolayer graphene \cite{Geim:07:1}. Since the electromagnetic coupling constant in graphene is effectively enhanced by the factor $c/v_F \approx 300$, where c is the velocity of light and $v_F$ is Fermi velocity, charge carriers turn out to be strongly coupled, and various non-perturbative phenomena such as spontaneous breaking of chiral (sublattice) symmetry can emerge. %PB
The existence of an insulating phase associated with chiral symmetry breaking is one of the central questions for the theory of graphene. Since analytic calculations become in general unreliable in the vicinity of the phase transition, the position of the transition point can only be found from first-principle numerical simulations. %PB

 %corr
%\textbf{ One of the central problems in this field is the question about possible semimetal-insulator phase transition in graphene due to appearance of chiral condensate. Since first-principle calculations can only be performed numerically in this situation,  various numerical calculations were performed to investigate this problem.}
 %corr

 In the seminal works \cite{Lahde:09:1} it has been realized that the low-energy effective theory of graphene at neutrality point can be efficiently simulated by the Hybrid Monte-Carlo method, which is commonly used in lattice Quantum Chromodynamics (QCD). In the more recent work \cite{Buividovich:12:1} Hybrid Monte-Carlo method was applied to perform a direct simulation of the tight-binding model of monolayer graphene (the possibility of such simulations was also discussed in \cite{Rebbi:11:1}). In these simulations only the nearest-neighbour hopping for the $\pi$ orbitals was considered, and inter-electron interactions were described by the Coulomb law (with some finite on-site interaction potential). So far all simulations, both with the low-energy effective theory and with the tight-binding model, have indicated that at the critical coupling constant $\alpha_c \approx 1$ there is a semimetal-insulator phase transition associated with the emergence of a mass gap in the quasiparticle spectrum due to spontaneous chiral symmetry breaking. According to these results suspended graphene, for which the effective coupling constant is $\alpha_s = e^2/ \hbar v_F \approx 300/137 \approx 2.2$, should be deeply in the insulating gapped phase with broken chiral symmetry (we note also that in this phase graphene is in fact anti-ferromagnetic \cite{Semenoff:11:1}).

 However, these findings are in clear contradiction with recent experimental studies of the Manchester group \cite{Elias:11:1}, in which no indications of the existence of a mass gap in suspended monolayer graphene were found. Till now the origin of this discrepancy between experimental and numerical data was not clear. In this paper we demonstrate that if one takes into account the screening of the Coulomb potential due to electrons on $\sigma$-orbitals of carbon, the interaction between electrons should be even stronger than in suspended graphene in order to trigger the semimetal-insulator phase transition. To this end we perform Hybrid Monte-Carlo simulations of the tight-binding model of monolayer graphene with the partially screened inter-electron interaction potential obtained in \cite{Wehling:11:1} in the constrained random phase approximation (cRPA). In the calculations of \cite{Wehling:11:1} only the screening due to $\sigma$-orbitals was taken into account, thus one can use it as an input to the tight-binding model of electrons on $\pi$-orbitals without any double-counting of screening terms.

 %%corr %TODO: insert this piece later in the text
%\textbf{In order to take into phenomenological potentials we sufficiently modified existing lattice Monte-Carlo methods. We used Hubbard-Stratonovich transformation instead of conventional lattice gauge field to introduce interaction term in fermionic action. Also some additional modifications have been done to suppress discretization errors.}
 %%corr

 The observed shift of the phase transition thus eliminates the controversy between experimental and numerical results and opens up the possibility of much more realistic first-principle Monte-Carlo simulations of the electronic properties of graphene. We further demonstrate that a rather mild increase of interaction strength do leads to spontaneous chiral symmetry breaking. Due to such proximity of the transition point, nonperturbative effects can be  quite important in suspended graphene.

 Since the screening of the Coulomb potential due to $\sigma$-orbitals is mostly important at small distances of the order of lattice spacing \cite{Wehling:11:1}, it seems that the position of the semimetal-insulator phase transition is highly sensitive to the form of the inter-electron interaction potential at short distances. We note that the high sensitivity of low-energy effective theory to ultraviolet regularization was also discovered in the work \cite{Zubkov:13:1}, where fermionic propagators were found to be saturated by momenta of the order of inverse lattice spacing.

 The fact that in suspended monolayer graphene the effective inter-electron interaction should be weaker than in the tight-binding model for the $\pi$ orbitals was also noted in \cite{Lahde:13:1} by fitting the numerical value of the renormalized Fermi velocity $v_F\lr{\alpha}$ to the experimental data of \cite{Elias:11:1}. The corresponding value of $\alpha$ was estimated as $\alpha \sim 0.7 \ldots 0.9$, which is significantly smaller than $\alpha_s$. %PB
Recent semi-analytic studies of the gap equations in graphene \cite{Popovici:13:1} has also shown that the phase transition is shifted to larger couplings if one takes into account the renormalization of the Fermi velocity. Our results provide a microscopic explanation of these findings. %PB

 The starting point of our simulations is the tight-binding Hamiltonian with the staggered potential $m$:
\begin{eqnarray}
\label{tbHam1}
 \hat{H}_{tb} = - \kappa \sum\limits_{<x,y>} \lr{ \hat{a}^{\dag}_{y} \hat{a}_{x}
+ \hat{b}^{\dag}_{y} \hat{b}_x + h.c.}
 + \nonumber \\ +
 \sum\limits_{x} \pm m \hat{a}^{\dag}_{x} \hat{a}_{x} \pm m \hat{b}^{\dag}_{x} \hat{b}_{x}.
\end{eqnarray}
where $\kappa = 2.7 \ev$, the sum $\sum\limits_{<x,y>}$ is performed over all pairs of nearest-neighbour sites of the graphene hexagonal lattice (we impose periodic spatial boundary conditions as in \cite{Buividovich:12:1}) and $\hat{a}^{\dag}$, $\hat{a}$ and $\hat{b}^{\dag}$, $\hat{b}$ are the creation/annihilation operators for particles and holes, respectively. The latter are related to creation/annihilation operators $\hat{c}^{\dag}_{x,s}$, $\hat{c}_{x,s}$ for electrons with spin $s = \uparrow, \downarrow$ as $\hat{a}_x = \hat{c}_{x, \uparrow}$, $\hat{b}_x = \pm \hat{c}^{\dag}_{x, \downarrow}$, where we take the plus sign for $x$ belonging to one of the simple sublattices of graphene hexagonal lattice and the minus sign - for another simple sublattice \cite{Rebbi:11:1, Buividovich:12:1}. The whole Hilbert space of the tight-binding model can be constructed by the action of the creation operators $\hat{a}^{\dag}_x$, $\hat{b}^{\dag}_x$ on the ground state $\ket{0}$ fixed by the conditions $\hat{a}_x \ket{0} = 0$, $\hat{b}_x \ket{0} = 0$. In this ground state each lattice site is occupied by one electron with spin down. Of course, in Monte-Carlo simulations we sum over all possible states of the system, so this choice of the ground state is only motivated by calculational convenience.

 The staggered potential is equal to $+m$ for the sites of one simple sublattice and $-m$ for sites of another simple sublattice. Its role is twofold: first, it regularizes the inverse of the fermionic kinetic operator in the Hybrid Monte-Carlo algorithm \cite{Buividovich:12:1, Rebbi:11:1}. Second, the staggered potential explicitly breaks the chiral (sublattice) symmetry and thus serves as a seed for spontaneous chiral symmetry breaking, which would otherwise be impossible in a finite volume. In the low-energy effective theory $m$ corresponds to the Dirac mass.

 Next we introduce the interaction Hamiltonian with an inter-electron interaction potential $V_{xy}$:
\begin{eqnarray}
\label{cHam}
 \hat{H}_{C} = {1 \over 2} \, \sum\limits_{x,y} V_{xy} \hat{q}_x \hat{q}_y,
\end{eqnarray}
where $\hat{q}_x = \hat{a}^{\dag}_{x} \hat{a}_{x} - \hat{b}^{\dag}_{x} \hat{b}_{x}$ is the operator of electric charge at lattice site $x$.

\begin{figure}
 \centering
 \includegraphics[width=7.5cm]{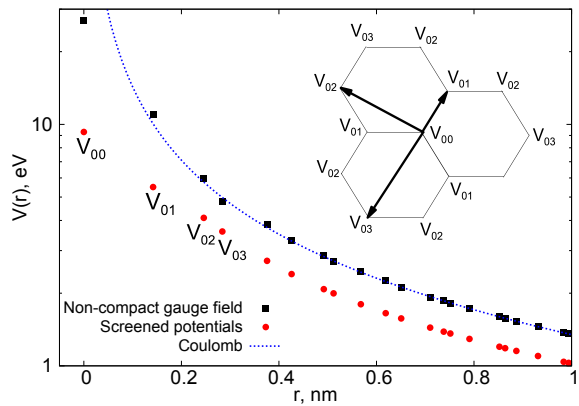}\\
 \caption{A comparison of the partially screened Coulomb potential with the exact Coulomb potential and the potential obtained from non-compact gauge field on the hexagonal lattice \cite{Buividovich:12:1}.}
 \label{fig:potentials}
\end{figure}

 For the on-site interaction potential $V_{xx} \equiv V_{00}$ and the potentials between nearest ($V_{01}$), next-to-nearest ($V_{02}$) and next-to-next-to-nearest-neighbouring lattice sites ($V_{03}$) we use the values calculated in \cite{Wehling:11:1} (see table I, 3d column). The resulting shape of the potential is illustrated on Fig. \ref{fig:potentials}. At larger distances we use the Coulomb potential $V\lr{r} = 1/\lr{\epsilon_{\sigma} r}$. The form of the potential is additionally corrected to account for periodic boundary conditions. The factor $\epsilon_{\sigma} \approx 1.41$ describes the contribution of electrons on $\sigma$ orbitals to the effective dielectric permittivity of graphene monolayer at intermediate distances and is obtained by equating $V_{03}$ to the Coulomb potential at $r = r_{03} = 0.284 \nm$: $V_{03} = 1/\lr{\epsilon_{\sigma} r_{03}}$. Physically this means that we assume that all the charges which screen the potential of a test charge are localized within the radius $r_{03}$. It is important to stress that this large-distance correction of the potential by a factor $1/\epsilon_{\sigma}$ alone is insufficient to prevent the semimetal-insulator phase transition in suspended graphene. Indeed, since for the unscreened Coulomb potential the corresponding critical value of the coupling constant $\alpha_c \approx 1$ \cite{Lahde:09:1, Buividovich:12:1} is more than two times smaller than the effective coupling constant $\alpha_s \approx 2.2$ in suspended graphene, the coefficient $\epsilon_{\sigma}$ should be at least larger than $2$ in order to shift the phase transition to $\alpha > \alpha_s$. Since two-dimensional fermions cannot screen the three-dimensional Coulomb potential at asymptotically large distances, in this limit $V\lr{r}$ should approach the unscreened Coulomb potential $V\lr{r} = 1/r$. However, with lattice sizes which we use in our simulation this asymptotics is in fact not yet relevant.

 We proceed by making the standard Suzuki-Trotter decomposition of the partition function:
\begin{eqnarray}
\label{SuzukiTrotter}
 \tr e^{- \beta \lr{\hat{H}_{tb} + \hat{H}_C} }  = \tr \lr{e^{- \delta \lr{\hat{H}_{tb} + \hat{H}_C} }}^{N_t} = \nonumber \\ =
\tr\lr{e^{-\hat{H}_{tb} \delta} e^{-\hat{H}_{C} \delta} e^{-\hat{H}_{tb} \delta} \ldots } + O\lr{\delta^2},
\end{eqnarray}
where $\beta = \lr{k T}^{-1}$ is the inverse temperature and $\delta = \beta/N_t$ with $N_t \gg 1$. The factors in the last line of (\ref{SuzukiTrotter}) are now interleaved with decompositions of the identity operator over Grassmann coherent states:
\begin{eqnarray}
\label{I_represent}
I = \int d\psi d\eta d\bar{\psi} d\bar{\eta} \, e^{-\sum\limits_x \bar{\psi}_x \psi_x - \sum\limits_x \bar{\eta}_x \eta_x} \,
\ket{\psi, \eta} \bra{\psi, \eta},
\nonumber \\
\ket{\psi, \eta} = e^{-\sum\limits_x \psi_x \hat{a}^{\dag}_x + \eta_x \hat{b}^{\dag}_x} \ket{0} .
\end{eqnarray}
The matrix elements $\bra{\psi, \eta} e^{-\delta \hat{H}_{tb}} \ket{\psi', \eta'}$ can be now easily calculated using the identity
\begin{eqnarray}
 \label{matrix_el}
 \bra{\psi} e^{\sum\limits_{x,y} \hat{a}^{\dag}_x A_{x y} \hat{a}_y } \ket{\psi'}
 = %\nonumber \\ =
 \expa{ \sum\limits_{x,y} \bar{\psi}_{x} \lr{e^A}_{xy} \psi_{y}' }  .
\end{eqnarray}
In order to find the matrix elements of the exponent of the interaction Hamiltonian $\hat{H}_C$ we perform the Hubbard-Stratonovich transformation \cite{Rebbi:11:1}:
\begin{eqnarray}
\label{HubbStr}
\expa{-{\delta \over 2} \sum\limits_{x,y} \hat{q}_x V_{xy} \hat{q}_y}
\cong \nonumber \\ \cong
\int\mathcal{D}\varphi_x \expa{ -{\delta \over 2} \sum\limits_{x,y} \varphi_x
V^{-1}_{xy} \varphi_y - i \delta \sum\limits_{x} \varphi_x \hat{q}_x }  ,
\end{eqnarray}
where $V^{-1}_{xy}$ is the matrix inverse of the potential $V_{xy}$: $\sum\limits_{z} V^{-1}_{xz} V_{zy} = \delta_{xy}$. After that we again apply the formula (\ref{matrix_el}) to the last line of (\ref{HubbStr}) and finally arrive at the following functional integral representation of the partition function:
\begin{eqnarray}
\label{PartFunc1}
\tr e^{-\beta \hat{H}} =
\int \mathcal{D}\varphi_{x,n} \mathcal{D}\psi_{x,n} \mathcal{D}\eta_{x,n} \mathcal{D}\bar{\psi}_{x,n}
\mathcal{D}\eta_{x,n}
\nonumber \\
e^{ - S\lrs{\varphi_{x, n}} - \sum\limits_{x,y,n,n'} \lr{\bar{\eta}_{x,n} \bar{M}_{x,y,n,n'} \eta_{y,n'} +
         \bar{\psi}_{x,n} M_{x,y,n,n'} \psi_{y,n'}}  },
\end{eqnarray}
where $S\lrs{\varphi_{x, n}} ={\delta \over 2} \sum\limits_{x,y,n} \varphi_{x,n} V^{-1}_{xy} \varphi_{y,n}$ is the action of the Hubbard field $\varphi_{x, n}$ and $n = 0~\ldots~{2N_t-1}$ enumerates the factors in the last line of (\ref{SuzukiTrotter}). The fermionic part of the action is written as follows:
\begin{eqnarray}
\label{ferm_action}
 \sum\limits_{x,y,n,n'} \bar{\psi}_{x,n} M_{x,y,n,n'} \psi_{y,n'}
 = \nonumber \\ =
 \sum\limits_{k=0}^{N_t-1}
 \left[ \sum\limits_x \bar{\psi}_{x,2k} \lr{\psi_{x,2k}-\psi_{x,2k+1}} \right.
 \nonumber \\
 -\delta \, \kappa \sum\limits_{<x,y>}
 \lr{ \bar{\psi}_{x,2k} \psi_{y,2k+1} + \bar{\psi}_{y,2k} \psi_{x,2k+1}}
 \nonumber \\
 + \sum\limits_x \bar{\psi}_{x,2k+1}\lr{\psi_{x,2k+1}  - e^{-i\delta \, \phi_{x,k}} \psi_{x,2k+2}}
 + \nonumber \\ +
 \left. \delta \sum\limits_x \pm m \bar{\psi}_{x,2k} \psi_{x,2k+1} \right] .
\end{eqnarray}
In this expression the Grassmann variables $\psi_{x,2k}$ and $\psi_{x,2k+1}$ label the fermionic coherent states inserted between the factors $e^{-\hat{H}_{tb} \delta}$, $e^{-\hat{H}_{C} \delta}$ and $e^{-\hat{H}_{C} \delta}$, $e^{-\hat{H}_{tb} \delta}$ in (\ref{SuzukiTrotter}), respectively. It can be shown that such a ``double-layer'' structure of the action leads to discretization errors of the order of $\delta$, in contrast to simpler fermionic action constructed in \cite{Rebbi:11:1}, for which discretization errors scale as $\sqrt{\delta}$. In practice, this form of the action allows one to obtain numerical results with sufficiently good precision even at quite coarse lattices ($N_t \sim 10 \ldots 20$, $\delta \sim 0.1 \, \kappa$). We also impose anti-periodic boundary conditions in time direction on fermionic variables $\psi_{x,n}$, $\eta_{x,n}$ in (\ref{ferm_action}).

 Now the Grassmann variables in (\ref{PartFunc1}) can be integrated out, which yields the following representation of the partition function:
\begin{eqnarray}
\label{PartFunc2}
 \tr e^{-\beta \hat{H}} \cong \int \mathcal{D}\varphi_{x,n} e^{-S\lrs{\varphi_{x,n}}}
 |\det{M\lrs{\varphi_{x,n}}}|^2 .
\end{eqnarray}
The manifest positivity of the integration weight in (\ref{PartFunc2}) is due to the symmetry between particles and holes for graphene at neutrality point. For example, at finite chemical potential the two fermionic determinants appearing in (\ref{PartFunc2}) after integration over $\psi_{x, n}$ and $\eta_{x,n}$ in (\ref{PartFunc1}) would no longer be complex conjugate, which would make Monte-Carlo simulations much more difficult due to the fermionic sign problem. For our choice of the inter-electron interaction potential, the action of the Hubbard field $S\lrs{\varphi_{x,n}}$ is also a positive definite quadratic form. Thus we can generate the configurations of $\varphi_{x,n}$ by a Monte-Carlo method and calculate physical observables as averages over the generated configurations. Here we follow \cite{Buividovich:12:1, Rebbi:11:1} and use the Hybrid Monte-Carlo method with the $\Phi$-algorithm. Inversion of the fermionic operator $M$, which is the most difficult part of this algorithm, was accelerated using GPUs.

 In order to detect the chiral symmetry breaking, we calculate the chiral condensate, which is the difference of particle numbers on the two simple sublattices $A$ and $B$:
\begin{eqnarray}
\label{observable}
 \vev{ \Delta n } ={1 \over N}
 \vev{
 \sum\limits_{x \in A} (\hat{a}^{\dag}_{x} \hat{a}_{x} +  \hat{b}^{\dag}_{x} \hat{b}_{x})
 -
 \sum\limits_{x \in B} (\hat{a}^{\dag}_{x} \hat{a}_{x} + \hat{b}^{\dag}_{x} \hat{b}_{x})
 } ,
\end{eqnarray}
where $N$ is the overall number of sites of one sublattice of hexagonal lattice. In terms of the fermionic operator $M_{x,y,n,n'}$ this expectation value reads:
\begin{eqnarray}
\label{observable1}
\vev{ \Delta n } ={1 \over {N N_t}} \sum\limits_{n=0}^{2 N_t-1}
\vev{\sum\limits_{x \in A} M^{-1}_{x,x,n,n}
-
\sum\limits_{x \in B} M^{-1}_{x,x,n,n}
} ,
\end{eqnarray}
where the average is now taken over configurations of the Hubbard field with the weight (\ref{PartFunc2}).

 Our simulations were performed on the lattice with spatial size $18 \times 18$ and $N_t = 20$, $\delta = 0.1 \ev^{-1}$, which corresponds to the temperature $T = 0.5 \ev = 5.8 \cdot 10^3 \K$. This temperature is considerably higher than in real experiments, however, in our simulations it is the temperature of the electron gas only. We do not consider thermal fluctuations of the crystalline lattice, thus phonon temperature is formally zero. We rely here on the results of \cite{Buividovich:12:1}, which indicate that as long as the electron temperature is much smaller than the hopping parameter $\kappa$ in (\ref{tbHam1}), it does not significantly affect the insulator-semimetal phase transition. To study the behavior of the condensate (\ref{observable}) in the limit $m \rightarrow 0$, we perform simulations at five different values of the staggered potential: $m = 0.1, \, 0.2, \, 0.3, \, 0.4, \, 0.5 \ev$. The interaction strength is controlled by additionally rescaling the potential by some factor $\epsilon$: $V_{xy} \rightarrow V_{xy} / \epsilon$.

 The coefficient $\epsilon$ can be thought of as the dielectric permittivity of the medium surrounding the graphene monolayer. However, to make this interpretation physically consistent one should also redo the calculations of \cite{Wehling:11:1} taking into account this additional screening. In our case $\epsilon$ has no direct physical interpretation and is only used to characterize the proximity of suspended graphene (which corresponds to $\epsilon = 1$) to the phase transition. For each set of lattice parameters we have generated $100$ statistically independent configurations of the field $\varphi_{x,n}$.

\begin{figure}
  \centering
  \includegraphics[width=7.5cm]{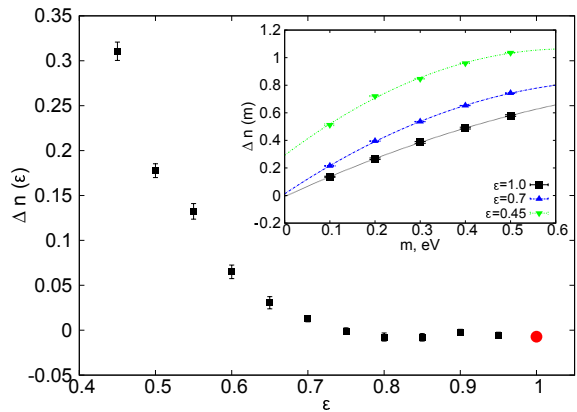}\\
  \caption{The dependence of the chiral condensate (\ref{observable1}) on $\epsilon$ and on $m$ (in the inset) for the $18 \times 18$ lattice with $N_t = 20$ and $\delta = 0.1 \ev^{-1}$. For $\epsilon = 1.0$ we show the results obtained on the $24 \times 24$ lattice with $N_t = 40$, $\delta = 0.05 \ev^{-1}$.}
  \label{fig:condensate}
\end{figure}

 The dependence of the chiral condensate (\ref{observable1}) on $\epsilon$ for $\epsilon \le 1$ is illustrated on Fig. \ref{fig:condensate}. To obtain the plotted values of $\Delta n$, we have fitted the mass dependence of the condensate $\Delta n\lr{m}$ by a quadratic function of $m$ and used this fit to extrapolate $\Delta n\lr{m}$ to $m=0$. These fits are shown on Fig. \ref{fig:condensate} in the inset. One can see that the extrapolated value $\Delta n\lr{m \rightarrow 0}$ for suspended graphene ($\epsilon = 1$) is equal to zero within error range, which indicates the absence of chiral symmetry breaking. We have also checked this result on the larger ($24 \times 24$, $N_t = 20$, $\delta = 0.1 \ev^{-1}$) and finer ($24 \times 24$, $N_t = 40$, $\delta = 0.05 \ev^{-1}$) lattices and on the larger set of 250 configurations of $\varphi_{x, n}$. All our measurements confirm that after extrapolation to $m = 0$ the chiral condensate is equal to zero for suspended graphene.

 Only at $\epsilon < \epsilon_c \approx 0.7$ the extrapolation to $m \rightarrow 0$ yields nonzero chiral condensate, which suggests that the state with broken chiral symmetry is favoured, and spontaneous chiral symmetry breaking is likely in the infinite volume limit. The fact that the critical value $\epsilon_c \approx 0.7$ is quite close to one suggests that while suspended graphene is still in the conducting phase with unbroken chiral symmetry, the proximity of the phase transition can still manifest itself in large fluctuations of order parameter (chiral condensate) and in other non-perturbative phenomena.

 We conclude that the screening of the Coulomb potential by electrons on $\sigma$-orbitals strongly influences the insulator-semimetal phase transition in monolayer graphene, so that the transition point is shifted into the region of parameter space in which the interaction strength is even stronger than in suspended graphene. This shift provides possible explanation of the long standing discrepancy between numerical \cite{Lahde:09:1, Buividovich:12:1} and experimental \cite{Elias:11:1} data on spontaneous gap generation in suspended graphene. We also note an intriguing possibility to effectively enhance the inter-electron interactions by stretching the graphene layer \cite{Wehling:11:1}, which can be used to reach the transition point in experiment.

\acknowledgments We thank V.V.~Braguta, T.~L\"{a}hde, L.~von~Smekal and D.~Smith for useful discussions. The work of the Moscow group was supported by Grants RFBR-11-02-01227-a, RFBR-12-02-31249 and RFBR-13-02-01387, Federal Special-Purpose Program "Cadres" of the Russian Ministry of Science and Education,
and by a grant from the FAIR-Russia Research Center. Numerical calculations were performed at the ITEP computer systems ``Graphyn'' and ``Stakan'' and at the supercomputer center of Moscow State University.
The work of P.B. was supported by the S.~Kowalewskaja award from the Alexander von Humboldt foundation (sponsored by the Ministry for Education and Research of the German Federal Republic).  MIK acknowledges a support form Stichting Fundamenteel Onderzoek der Materie (FOM), The Netherlands.

%\bibliography{MyBibliography}
%\bibliographystyle{apsrev}

\end{document}